\documentclass[secnumarabic, amssymb, nobibnotes, aps, prb, reprint, superscriptaddress]{revtex4-2}
\usepackage[utf8]{inputenc}
\usepackage[english]{babel} %uncomment if necessary
\usepackage{amsmath}
\usepackage{amssymb}
\usepackage{upgreek}
\usepackage{gensymb} % Für \degrees
\usepackage{xfrac} % Bessere Brüche

\usepackage{graphicx}
\usepackage{float}
\usepackage[font=small, justification=raggedright, format=plain]{caption} % justification=justify

\usepackage{hyperref}
\begin{document}
	
	\title{Ultrafast transport mediated homogenization of photoexcited electrons governs the softening of the $A_\mathrm{1g}$ phonon in bismuth}%
	
	\author{F.~Thiemann}%
	\email[]{fabian.thiemann@uni-due.de}
	\affiliation{Department of Physics, University of Duisburg-Essen, Lotharstrasse 1, 47057 Duisburg, Germany}
	\author{G.~Sciaini}
	\affiliation{The Ultrafast Electron Imaging Lab, Department of Chemistry, and Waterloo Institute for Nanotechnology, University of Waterloo, Waterloo, Ontario N2L 3G1, Canada}
	\author{A.~Kassen}
	\affiliation{Department of Physics, University of Duisburg-Essen, Lotharstrasse 1, 47057 Duisburg, Germany}
	\author{U.~Hagemann}
	\affiliation{Interdisciplinary Center for Analytics on the Nanoscale (ICAN), University of Duisburg-Essen, Carl-Benz-Straße 199, 47057, Duisburg, Germany}
	\affiliation{Center for Nanointegration (CENIDE), University of Duisburg-Essen, Carl-Benz-Str. 199, 47057 Duisburg, Germany}
	\author{F.~\surname{Meyer zu Heringdorf}}
	\affiliation{Department of Physics, University of Duisburg-Essen, Lotharstrasse 1, 47057 Duisburg, Germany}
	\affiliation{Interdisciplinary Center for Analytics on the Nanoscale (ICAN), University of Duisburg-Essen, Carl-Benz-Straße 199, 47057, Duisburg, Germany}
	\affiliation{Center for Nanointegration (CENIDE), University of Duisburg-Essen, Carl-Benz-Str. 199, 47057 Duisburg, Germany}
	\author{M.~\surname{Horn-von Hoegen}}
	\affiliation{Department of Physics, University of Duisburg-Essen, Lotharstrasse 1, 47057 Duisburg, Germany}
	\affiliation{Center for Nanointegration (CENIDE), University of Duisburg-Essen, Carl-Benz-Str. 199, 47057 Duisburg, Germany}
	
	\date{May 2022}%
	
	\begin{abstract}
	    In order to determine the role of non-thermal transport of hot carriers which is decisive for the dissipation of energy in condensed matter we performed time-resolved broadband femtosecond transient reflectivity measurements on $7$~–~$197\,\mathrm{nm}$ thick Bi(111) films epitaxially grown on Si(111). We monitored the behavior of the Fourier amplitude and the central frequency of the coherent $A_\mathrm{1g}$ phonon mode as function of the incident fluence, film thickness, and probe wavelength in the range of $580$~\nobreakdash-~$700\,\mathrm{nm}$. The frequency redshift that follows photoexcitation was used as a robust quantity to determine the effective distribution of excited carriers that governs the displacive excitation mechanism of coherent $A_\mathrm{1g}$ phonons in Bi. For Bi films up to $50\,\mathrm{nm}$ thickness a homogeneous excitation due to the ultrafast transport of hot charge carriers is observed, limited by a carrier penetration depth of $60\,\mathrm{nm}$ independent of the totally deposited laser energy.
	\end{abstract}
	\maketitle
	
	%\tableofcontents % Do no use in this paper.
	
	%%% INTRO %%%
	Bismuth as a Peierls-Jones distorted semimetal~\cite{Peierls1991} with its low equilibrium carrier density and its low effective masses is prone to carrier induced structural instabilities. One of them is the distinct displacive excitation of coherent $A_\mathrm{1g}$ optical phonons which has turned bismuth into one of the most studied materials by ultrafast time-resolved pump-probe techniques~\cite{Cheng1990, Hase1996, DeCamp2001, Boschetto2008, Shin2015, Ishioka2015, Teitelbaum2018, SokolowskiTinten2003, Fritz2007, Sciaini2009, Johnson2009, SokolowskiTinten2015, Qi2020}. Upon photoexcitation the charge carrier density can be transiently increased by orders of magnitude thus drastically changing the potential energy surface of the atoms~\cite{SokolowskiTinten2003, Murray2005, Fritz2007}. With growing density of excited carriers the displacement amplitude of the $A_\mathrm{1g}$ phonon increases \cite{Murray2005, Fritz2007} as the mode significantly softens \cite{DeCamp2001, Hase2002}. In turn, the atomic displacement through the $A_\mathrm{1g}$ phonon strongly modulates the dielectric properties~\cite{Stevens1999} of the material enabling its observation via femtosecond transient reflectivity (fs-TR) measurements~\cite{Cheng1990, Hase1996, DeCamp2001, Boschetto2008, Shin2015, Ishioka2015, Teitelbaum2018}. This unique correlation between the density of photoexcited electrons and the behavior of the coherent $A_\mathrm{1g}$ phonon mode provides a quantity to explore the ultrafast spatial redistribution of excited carriers.

	Most fs-TR studies of Bi have focused on the investigation of the dynamics of coherent $A_\mathrm{1g}$ phonons in either bulk single crystalline samples or relatively thick Bi films of more than $100\,\mathrm{nm}$ at a wide range of excitation conditions~\cite{DeCamp2001, Hase2002, Garl2008, Shin2015, Teitelbaum2018} and temperatures~\cite{Hase1998, Ishioka2006, Garl2008}. Recent attention has been attracted by the study of coherent $A_\mathrm{1g}$ phonons in ultrathin films~\cite{Ishioka2015, Shin2015, He2020, Jnawali2021}. Among them  Shin~\emph{et~al}.~\cite{Shin2015} performed detailed fs-TR experiments on Bi films with thicknesses down to $25\,\mathrm{nm}$ on glass and Si substrates. They concluded that confinement yields a significantly higher density of excited carriers, resulting in increased mode softening and faster dephasing. Studies on bulk samples by Johnson~\emph{et~al}.~\cite{Johnson2008} and Boschetto~\emph{et~al}.~\cite{Boschetto2010} already indicated the existence of hot carrier transport that later became relevant in thin films~\cite{Jnawali2021}. Moreover, in a very recent study of Bi films grown on NaCl(001) single crystalline substrates, Jnawali~\emph{et~al}.~\cite{Jnawali2021} were able to explore the effects of hot carrier transport and empirically determine the effective hot carrier penetration depth that governs the amplitude of the coherent $A_\mathrm{1g}$ phonon. However,  their use of a single color probe in the limit of low carrier excitation limited the general applicability of this carrier transport model.
	
	Within fs-TR techniques, the flexibility to control the pump and probe central wavelengths has been essential to provide detailed insights into the nature of the generation mechanism of $A_\mathrm{1}$ phonons~\cite{Stevens1999, Stevens2002} and reveal the role of stimulated Raman scattering in such a process. A step ahead in spectral resolved probing is the implementation of ultrashort broadband (or supercontinuum) white light pulses. They enable broad spectral detection and the observation of spectroscopic effects that are not evident when single or two-color fs-TR are employed due to e.g. thin film interference effects or absorption minima and maxima arising from critical points in the dielectric function~\cite{Cheng2022}.

	\begin{figure*}[t!]
		\centering\includegraphics{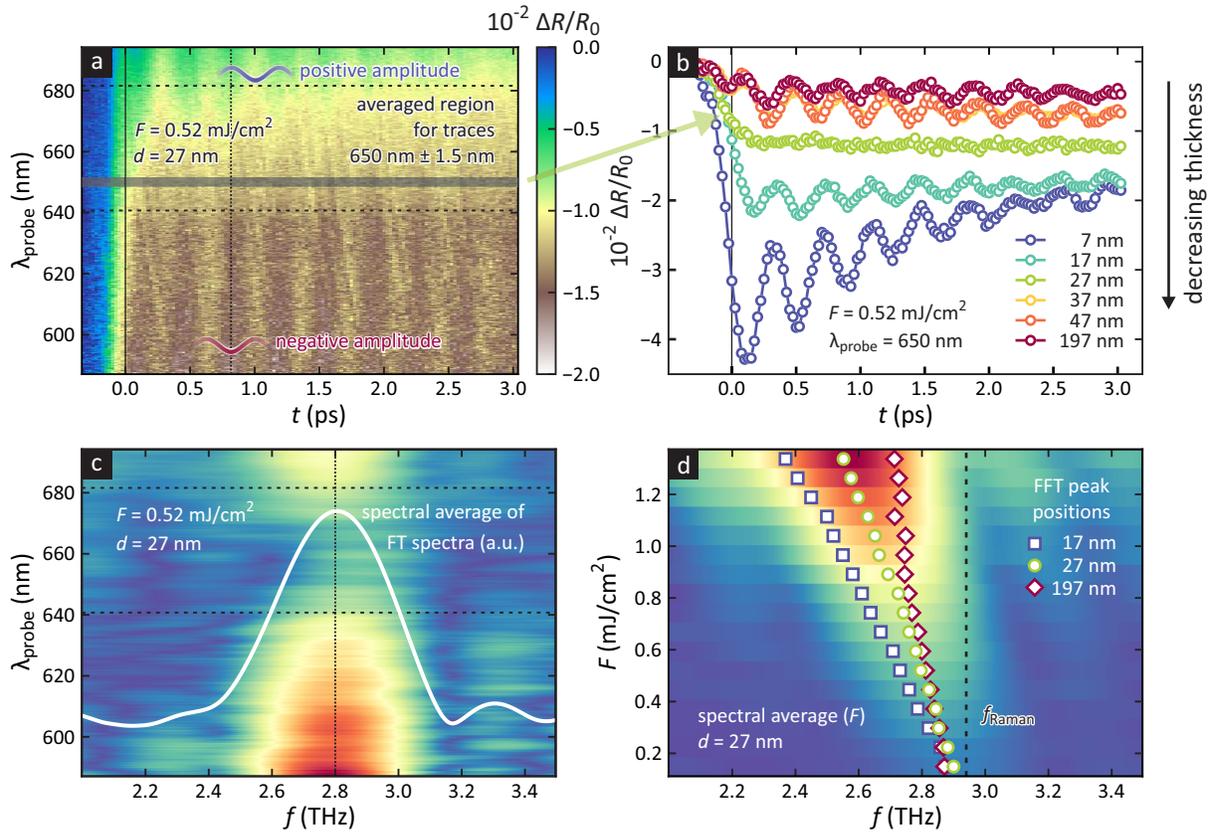}
		\caption{
			(Color online) (a) time-resolved bb-fs-TR spectrum obtained for the $27\,\mathrm{nm}$ thick Bi film. The sinusoidal traces are a guide for the eye. (b) Temporal traces obtained for all films at $\lambda_{\mathrm{probe}} = 650$ nm after averaging along a bin of 3 nm in width as depicted in (a) for $d =$ 27 nm. (c) Fourier power spectrum obtained from the analysis of the tr-bb-TA spectrum shown in (a). The white trace represents the spectral average. (d) Spectral averages of the Fourier power spectra for $d=27\,\mathrm{nm}$ and the peak value for $d=17\,\mathrm{nm}$, $d=27\,\mathrm{nm}$ and $d=197\,\mathrm{nm}$ as a function of $F$ compared to the Raman frequency (dashed line) of the $197\,\mathrm{nm}$ thick film, serving as a bulk reference.
		}\label{fig:data}
	\end{figure*}

	In the present work we implemented broadband fs-TR (bb-fs-TR) to study the dynamical behavior of coherent $A_\mathrm{1g}$ phonons in single crystalline Bi films of different thicknesses. Our experiments reveal a rather complex spectroscopic behavior in the bb-fs-TR spectra as function of film thickness ($d$) and incident pump fluence ($F$). A detailed analysis of the redshift ($\Delta f_c$) of the $A_\mathrm{1g}$ phonon central frequency ($f_c$) was performed. The results show that $\Delta f_c$ is the ideal quantity to determine the degree of excitation and the distribution of excited carriers. The effective hot carrier penetration depth ($d_{\mathrm{eff}}$) was estimated from measurements in the thin film limit (${d<d}_{\mathrm{eff}}$) where the density of excited carriers $n$ can be considered constant across the entire film. We observe that hot carrier transport happens within $150\,\mathrm{fs}$, which corresponds to half an oscillation period of the $A_\mathrm{1g}$ phonon and could be attributed to either ballistic or diffusive processes~\cite{Hayes1986, Gusev1998, Wright2001, Hada2012, Block2019, Sung2019}.
	
	%%% EXPERIMENTAL %%%
	Our bb-fs-TR setup is driven by an amplified Ti:Sapphire laser with a pulse energy of $1\,\mathrm{mJ}$ and a repetition rate of $5\,\mathrm{kHz}$ that delivers $160\,\mathrm{fs}$ pump pulses centered at a photon wavelength of $800\,\mathrm{nm}$. Relative transient reflectivity changes ($\Delta R/R_0$) originating from variations in electron and hole populations, hot-carrier energy relaxation processes, and coherent phonon oscillations were probed by time-delayed supercontinuum white light pulses and detected in a synchronized dispersive spectrometer, which provided a spectrum that spanned from $\lambda_{\mathrm{probe}}\approx 580\,\mathrm{nm}$ to $700\,\mathrm{nm}$. The principle of our setup follows those of others employing the well established bb-fs-TR method \cite{Kovalenko1999}. The supercontinuum was generated by focusing a small fraction of the fundamental beam power ($<1\,\mathrm{\mu J}$) into a $4\,\mathrm{mm}$ thick YAG crystal. The probed spectral window was limited in the short wavelength region by the gradually decreasing signal to noise ratio, and in the long wavelength limit by a $700\,\mathrm{nm}$ shortpass filter placed in front of the spectrometer to block the scattered pump light. A non-collinear arrangement with an angle of $30^\circ$ between the incident pump and probe beams was used. The full width at half maximum (fwhm) of the pump- and probe beam spot sizes at the film’s surface were approximately $180\,\mathrm{\mu m}$ and $130\,\mathrm{\mu m}$, respectively. The incident pump fluence was varied through the combination of a half-wave plate and a calcite polarizer. The time-resolved bb-fs-TR spectra ($\Delta R/R_0 \left(t, \lambda_\mathrm{probe}\right)$) were recorded by periodically blocking the pump beam with a mechanical chopper at $25\,\mathrm{Hz}$ and through an optical time delay. The typical time step was $25\,\mathrm{fs}$.

	[111]\nobreakdash-oriented epitaxial Bi films of various nominal thicknesses ($10\,\mathrm{nm}$, $20\,\mathrm{nm}$, $30\,\mathrm{nm}$, $40\,\mathrm{nm}$, $50\,\mathrm{nm}$, and $200\,\mathrm{nm}$) were grown on Si(111) substrates as described in references~\cite{Kammler2005, Payer2012} using a commercial electron beam evaporator. The crystallinity of the films was verified \emph{in situ} using LEED. The Bi film thickness was monitored during deposition using a quartz balance, that was calibrated by profilometry and atomic force microscopy to the thickness of reference samples. The equilibrium $A_{1g}$ frequencies in the films were measured employing Raman spectroscopy with $532$ and $633\,\mathrm{nm}$ exciting wavelength. Due to the exposure to ambient conditions the nominal values of the film thickness were corrected by subtracting the approximately $3\,\mathrm{nm}$ thick layer of $\mathrm{Bi_2 O_3}$~\cite{Payer2012}. Therefore, the effective Bi thicknesses $d$ in our experiments were $(7\pm1)\,\mathrm{nm}$, $(17\pm2)\,\mathrm{nm}$, $(27\pm3)\,\mathrm{nm}$, $(37\pm3)\,\mathrm{nm}$, $(47\pm4)\,\mathrm{nm}$, and $(197\pm11)\,\mathrm{nm}$.

	%%% RESULTS %%%
	The $\Delta R/R_0(t)$ spectrum obtained for the $27\,\mathrm{nm}$ thick Bi film at $F=0.52\,\mathrm{mJ/cm^2}$ is depicted in Fig.~\ref{fig:data}~(a). Note that in the range of $\lambda_{\mathrm{probe}}\approx640\,\mathrm{nm}$ to $680\,\mathrm{nm}$ (between the dashed lines), the amplitude of the oscillation flips its sign, i.e., suffers a relative $\pi$-phase shift. The two sinusoidal traces  in Fig.~\ref{fig:data}~(a) are placed as a guide for the eye and illustrate that the initial oscillation amplitude is negative below $640\,\mathrm{nm}$ and positive above $680\,\mathrm{nm}$. The marked area at $\left(650.0 \pm 1.5 \right)\,\mathrm{nm}$ (ten pixels in the spectrometer) in (a) illustrates the spectral range of averaging for the traces shown in Fig.~\ref{fig:data}~(b). Ultimately the sign flip of $\Delta R/R_0$ in (a) leads to a phase node or non-oscillatory signal in the time domain at $\lambda_{\mathrm{probe}}\approx650\,\mathrm{nm}$ as seen in the corresponding trace for $d=27\,\mathrm{nm}$ in (b). This renders any analysis based on the direct comparison of phonon oscillation amplitudes at a single probe wavelength near $\lambda_{\mathrm{probe}}\approx 650\,\mathrm{nm}$ very challenging, highlighting the advantages of broadband- over single-color probes~\cite{Cheng2022}.
	
\begin{figure*}[t!]
		\centering\includegraphics{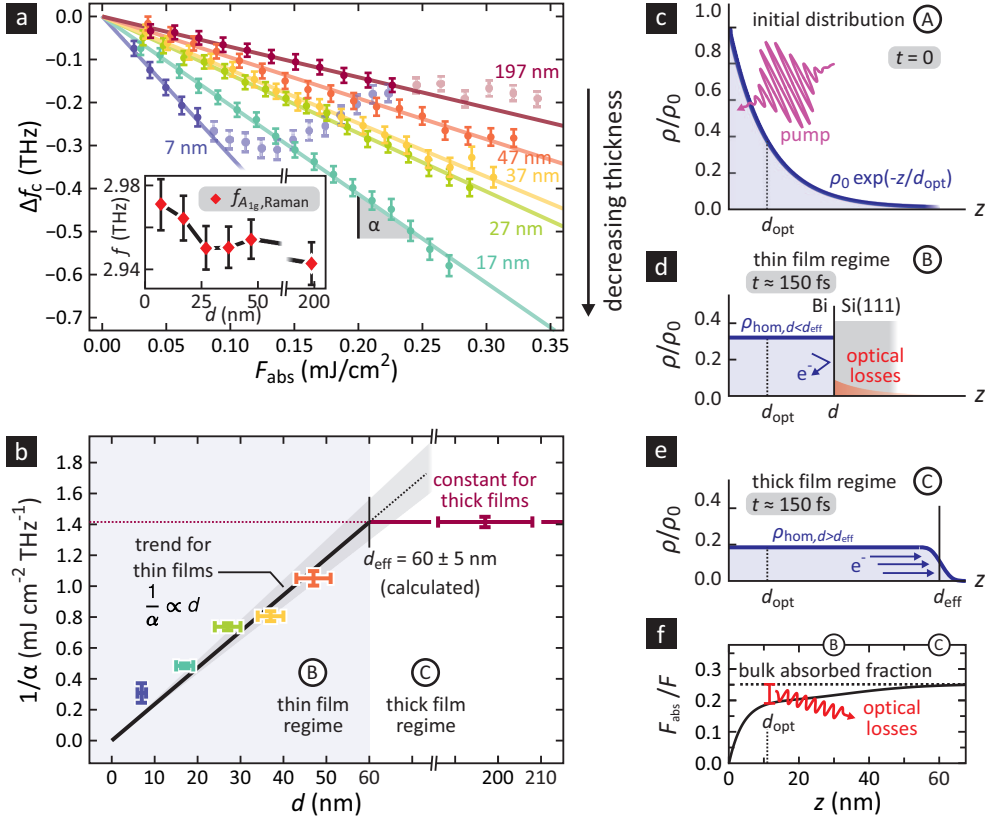}
		\caption{
			(Color online) (a) Frequency shift $\mathrm{\Delta}f_c=f_c-f_{0}$ of the $A_\mathrm{1g}$ mode as a function of $F_\mathrm{abs}$ for films of different thickness. The solid lines represent the slopes of the linear fit $f_c=f_{0}-\alpha F_\mathrm{abs}$. The inset shows the equilibrium frequency of the $A_\mathrm{1g}$ mode $f_\mathrm{Raman}\approx f_0$ obtained from Raman spectroscopy as a function of film thickness $d$. (b) Inverse slope $1/\alpha$ from (a) as a function of $d$. The thin film regime (B) is represented by a linear trend (black line) and the $d=197\,\mathrm{nm}$ value serves as a bulk reference  (horizontal line) in the thick film regime (C). The intersection of both determines the effective hot carrier penetration depth $d_\mathrm{eff}=60\,\mathrm{nm}$. The standard deviation of the trend's linear fit is marked by the gray area. (c) Distribution of energy in the carrier system within the first few femtoseconds (A) after excitation, represented by the absorbed energy density $\rho(z)=\rho_0 \exp\left(z/d_\mathrm{opt}\right)$ whereas $d_\mathrm{opt}$ is the calculated optical penetration depth. (d) and (e) show the homogeneous distribution of $\rho(z)$ for the thin- (B) and thick (C) film regime at later times. (f) Calculated ratio of absorbed ($F_\mathrm{abs}$) over incident fluence ($F$).
		}\label{fig:result}
	\end{figure*}

	Therefore, we employed fast Fourier transformation (FFT) in order to unambiguously obtain the frequency $f$, proportional to the degree of excitation, from the  $\Delta R/R_0(t)$ spectra. Figure~\ref{fig:data}~(c) shows the Fourier spectrum obtained  form the data shown in panel (a). We employed the time-derivative method to remove the slowly varying background signal arising from changes in the electronic population. This method provides the oscillatory residuals necessary to extract $f$ at each $\lambda_{\mathrm{probe}}$ via FFTs without the need to fit the temporal traces. The spectral region affected by the phase node $\lambda_{\mathrm{probe}}\approx 640\,\mathrm{nm}$ – $680\,\mathrm{nm}$ is now clearly visible as a zone where the FFT-amplitude of the coherent $A_\mathrm{1g}$ phonon is approximately zero. Note that the actual coherent $A_\mathrm{1g}$ phonon displacement modulates the dielectric properties of the material with the same frequency $f$ across the probed spectral window. Therefore, we averaged the FFT spectra along $\lambda_{\mathrm{probe}}$ to obtain the average FT spectrum (white trace) with the frequency at maximum $f_c\approx 2.8\,\mathrm{THz}$. Figure~\ref{fig:data}~(d) displays the average FT spectra for $d=27\,\mathrm{nm}$ as well as the values of $f_c$ obtained for $d=17\,\mathrm{nm}$, $d=27\,\mathrm{nm}$ and $d=197\,\mathrm{nm}$ as a function of $F$. The $f_c$ are compared to the measured Raman frequency ($f_{\mathrm{Raman}}$) (dashed line) for the thickest film, which can be considered as a bulk reference. $f_{\mathrm{Raman}}$ reflects the frequency $f_{0}$ of the $A_\mathrm{1g}$ mode without excitation.  This analysis procedure was carried out for all samples and the resulting frequencies $f_c$ and frequency changes $\Delta f_c = f_c - f_{0}$ are shown in Fig.~\ref{fig:data}~(d) and Fig.~\ref{fig:result}~(a) respectively.
	
	We observe an almost linear redshift in $f_c$ upon increasing $F$, as predicted by experiment~\cite{Hase2002} and theory~\cite{Murray2005}. Furthermore, the frequencies of the $A_\mathrm{1g}$ mode for films with different thicknesses $d$ shown in Fig.~\ref{fig:data}~(d) exhibit a pronounced difference in their fluence dependency. Since the frequency change $\Delta f_c$ depends on the number of excited carriers~\cite{Murray2005, Fritz2007}, we assumed that $\Delta f_c$ can be used as the quantity that determines the effective degree of excitation of the film. Therefore we relate it to the density of excited carriers $n(z)$ after hot carrier transport and thermalization or absorbed energy per unit volume $\rho(z)\propto n(z)$, that depends in general on the propagation distance from the film’s surface ($z$). More specifically, if it is assumed that carrier multiplication after absorption always results in the same number of excited carriers per photon, $\rho(z)$ should refer to the spatial distribution of the excited carriers. The term hot carrier density describes the initial distribution of charge carriers immediately after excitation whereas density of excited carriers is referred to as the distribution of carriers after transport and thermalization.
	
	%%% MODEL %%%
	The two extreme scenarios one would expect for $\rho(z)$ are (i) a complete homogeneous distribution of excited carriers ($n_\mathrm{hom}$) across the entire film thickness characterized by a constant absorbed energy density $\rho_\mathrm{hom}$ or (ii) a distribution of excited carriers that follows the spatial distribution of the originally absorbed pump light. The distribution of $\rho(z)$ normalized by the initially absorbed energy density $\rho_0$ at the surface is sketched for (ii) as case (A) in Fig.~\ref{fig:result}~(c) and for (i) as cases (B) and (C) in panel (d) and (e). The exemplary timestamps in (c-e) emphasize that (ii) is the initial distribution of hot carriers within the first few femtoseconds for cases (B) and (C) as well. In order to quantify the redshift $\Delta f_c$ and compare it with the scenarios (i) and (ii), we describe it with a linear slope ($\alpha$), and thus
	\begin{equation}
		\label{eq:fc}
		f_c(F_\mathrm{abs})=f_0-\Delta f_c = f_{0}-\alpha F_\mathrm{abs}.
	\end{equation}
	We fitted Eq. \eqref{eq:fc} to the measured $f_c$ for each film as shown by the solid lines in Fig.~\ref{fig:result}~(a). We found that $\alpha$ gradually increases with decreasing film thickness. The transparent data points shown in panel (a) are excluded from fitting due to radiation damage of the film ($d=7\,\mathrm{nm}$) and deviation from the trend with its origin yet unknown to us  ($d=197\,\mathrm{nm}$). Radiation damage was confirmed by rendering the recovery of low fluence results impossible after exceeding the threshold of $F_\mathrm{abs}\approx0.8\,\mathrm{mJ/cm^2}$. Here $f_{0}$ is the extrapolated value of $f_c$ at $F_\mathrm{abs}=0$ for a given $d$. Unfortunately, due to uncertainties in the fluence the error in $f_0$ is quite large. This is why we used the frequency shift $\Delta f_c$ and not the plain frequency $f_c$ as the relevant quantity. However, we still need to assure that differences in $f_0$ are significantly smaller than the observed redshift. Therefore we performed Raman spectroscopy to determine the frequency $f_0\approx f_\mathrm{Raman}$ of the $A_\mathrm{1g}$ mode without any carrier induced redshifts. $f_\mathrm{Raman}$ as a function of $d$ is shown in the inset in Fig.~\ref{fig:result}~(a). A clear blueshift in $f_\mathrm{Raman}$ towards thinner films is visible. This blueshift is likely induced by compressive strain, which is expected for Bi/Si(111) films, as studied by Payer~\emph{et~al.}~\cite{Payer2012}. Yet this blueshift is negligible compared to the pronounced redshift $\Delta f_c$. $F_\mathrm{abs}$ is the absorbed fluence that accounts for losses by reflection at the surface and transmission into the substrate as well as gains in absorption by reflection at the film/substrate interface, which all was calculated employing the transfer matrix method (TMM)~\cite{Prentice2000, Katsidis2002}. The ratio between $F_\mathrm{abs}$ and $F$ as a function of $d$ is shown in Fig.~\ref{fig:result}~(f). The optical losses at the interface are illustrated by the difference of $F_\mathrm{abs}(d)$ (solid line) and $\lim_{d\to\infty} F_\mathrm{abs}(d)$ (dotted line) in this panel. We safely assume that there are no losses of energy by carrier transport to the substrate due to the formation of a Schottky barrier at the Bi/Si(111) interface~\cite{Hricovini1992}. In the first case of a homogenous excited carrier distribution (i) $\Delta f_c$ as a function of $F_\mathrm{abs}$ follows
	\begin{equation}
		\label{eq:dfc}
		\Delta f_c \propto n_\mathrm{hom} \propto \rho_\mathrm{hom} = \frac{F_\mathrm{abs}}{d}.
	\end{equation}
	This inverse proportionality of $\Delta f_c$ and $\rho_\mathrm{hom}$ on $d$ results in the drastically increased slope~$\alpha$ (see Eq.~\ref{eq:fc}) towards thin films. The dramatic impact of inverse proportionality is shown in panels (d) and (e) of Fig.~\ref{fig:result}, comparing $\rho_\mathrm{hom}$ for two different film thicknesses excited with the same arbitrary $F$. The areas underneath the curves in (d) and (e) are approximately the same as in the initial distribution in (c) and reflects the totally absorbed energy in the film.

	We will show that in the thin film limit our observations can be well described by the first scenario (i), but we still need to rule out the second one (ii), especially because (ii) describes the initial distribution of hot carriers as well, before the carrier transport transforms the initial- into a homogeneous distribution of excited carriers (i). For the second scenario, i.e., ignoring hot carrier transport, carrier multiplication, and thus their effects on the $A_\mathrm{1g}$ vibrational coherence, $\rho(z)$ will follow the Beer-Lambert law for light absorption, $\rho (z)\nobreak=\nobreak\rho_0 \exp{\left(-z/d_\mathrm{opt}\right)}$ (see Fig.~\ref{fig:result}~(c)). For thin films we include corrections to account for the transmission and reflection of the pump field at the Bi/Si interface. We have calculated the optical penetration depth $d_\mathrm{opt}=11\,\mathrm{nm}$ for $\lambda =800\,\mathrm{nm}$ using the dielectric function of Bi measured by Toudert~\emph{et~al.}~\cite{Toudert2017} and found that thin film corrections to the value of $\rho_0$ are only within 8\% for the thinnest film. Considering that optical reflectivity essentially probes the surface of the film, the observed $A_\mathrm{1g}$ phonon frequency $f_c$ should be then proportional to the absorbed energy density at the surface $\rho_0$, which is still proportional to the fluence $F$, but nearly independent of the film thickness $d$. Thus, the expected values for $f_c(F)$ should be almost identical for all films. This is, however, not the case as shown by the varying redshifts for different $d$ in Fig.~\ref{fig:data}~(d). The redshift is about three times more pronounced in the $17\,\mathrm{nm}$ than in the $197\,\mathrm{nm}$ thick film. Therefore we render scenario (ii) as incorrect and continue with the first scenario (i). Previously we have described (i) as a homogeneous distribution $\rho(z)=\rho_\mathrm{hom}$ across the entire film thickness. Since bulk samples show demonstrably still significant redshifts in the $A_\mathrm{1g}$ mode, the transport of carriers has to stop at some distance while increasing the film thickness. To resolve this issue we introduce the thin- and thick film regimes. In the former the transport of carriers is limited by the film thickness $d$ (see Fig~\ref{fig:result}~(d)) and for the latter (see Fig~\ref{fig:result}~(e)) we introduce the limiting effective hot carrier penetration depth $d_\mathrm{eff}$. Note that the edge of $\rho(z)$ is sharp in (d) and smeared out in (e) due to the missing Schottky barrier in the thick film regime. In the thin film regime the combination of Eq.~\eqref{eq:fc} and Eq.~\eqref{eq:dfc} yields the following relationship, $1/\alpha \propto d$. As shown in Fig.~\ref{fig:result}~(b) the five thinnest samples exhibit the relation $1/\alpha \propto d$ and are within the thin film limit. We carried out a procedure similar to that presented by Jnawali~\emph{et~al}.~\cite{Jnawali2021} and determined $d_{\mathrm{eff}}=\left(60\pm5\right)\,\mathrm{nm}$ from the intersection of the thin film regime's linear trend (black line) with the bulk’s limit value (horizontal line, $d=197\,\mathrm{nm}$). This limit value cannot be exceeded, since bulk samples should exhibit the broadest spread of excited carriers.
	
	%%% DISCUSSION %%%
	We found that our $d_{\mathrm{eff}}$ is larger than the one found by Jnawali \emph{et~al.}~\cite{Jnawali2021}. A plausible explanation is a missing annealing step during growth of the Bi/NaCl(001) samples used by Jnawali~\emph{et~al.}, as described in their growth method originally published by Payer~\emph{et~al}.~\cite{Payer2008}. Annealing is a common method for defect reduction. Thus, in the case of Bi/NaCl(001) an increased amount of lattice defects remains in the Bi film. Since the electron mean free path is expected to depend on the sample quality (crystallinity, impurities, dislocations, etc.) the annealed films used in our study should exhibit a larger $d_{\mathrm{eff}}$. 
	
	Furthermore, we can conclude that the transport behavior in the thin film regime and effective hot carrier penetration depth $d_\mathrm{eff}$ is, within the measured range of fluences $F$, independent of the totally deposited energy. If the process of homogenization would be distorted, the ratios between $\Delta f_c$ for different $d$ would change as a function of $F_\mathrm{abs}$. For the behavior of $d_\mathrm{eff}$ two simple cases to discuss are a decreasing or increasing $d_\mathrm{eff}$ as a function of  $F_\mathrm{abs}$. The first results in a larger $|\Delta f_c|$ for $d=197\,\mathrm{nm}$ and eventually for $d=47\,\mathrm{nm}$ etc., whereas the latter results in a declining $|\Delta f_c|$ for $d=197\,\mathrm{nm}$. The latter case can be applied to some of the last six data points ($F_\mathrm{abs}>0.22\,\mathrm{mJ/cm^2}$) shown transparent in Fig.~\ref{fig:result}~(a). But since they scatter and show no clear trend, one has to be very careful concluding a modification of $d_\mathrm{eff}$. Therefore, hot carrier transport at higher excitation levels than the presented ones requires further attention.
	
	Concerning the nature of the hot carrier transport process, which may be diffusive as proposed by Johnson~\emph{et~al}.~\cite{Johnson2008}, ballistic as discussed by Brorson~\emph{et~al}.~\cite{Brorson1987} and Suárez~\emph{et~al}.~\cite{Suarez1995}, or likely a combination of both, we are unable to give an answer. Due to the temporal response function of our pump-probe setup ($\approx 160\,\mathrm{fs}$) we cannot differentiate between these two possible contributions. Such hot carrier transport mechanisms have been investigated recently by time-resolved microscopy in metals and semiconductors by Block~\emph{et~al}.~\cite{Block2019} and Sung~\emph{et~al}.~\cite{Sung2019} respectively. Those results show that not only the differences in propagation velocity are important, but that the time dependency of the mean square displacement of hot carriers is the deciding parameter for the transport mechanism. In our temporal traces the frequency redshifts are visible after half an oscillation period of the $A_\mathrm{1g}$ phonon’s mode. At this point, no changes in the instantaneous frequency, which would be an indication for a slower hot carrier transport, are observed. This was verified through time-windowed FT. We thus conclude that the spread of hot carriers occurs on a timescale shorter than $150\,\mathrm{fs}$, i.e., shorter than half an oscillation period of the $A_\mathrm{1g}$ vibrational coherence.
	
	In conclusion, we demonstrated a simple and improved approach to determine the effective hot carrier penetration depth, the excited carrier distribution and their fluence dependency in Bi, which is found to be independent of the totally deposited energy. Here, we employed the detected redshift of the $A_\mathrm{1g}$ mode to determine the absorbed energy density $\rho$ proportional to the density of excited carriers $n$. In contrast to the amplitude of the oscillatory component in $\Delta R/R_0\left(t\right)$ the change in frequency is not affected by thin film interference effects, absorption minima or maxima arising from critical points in the dielectric function. This allowed us to conclude that the hot carriers distribute homogeneously throughout thin films ($d<60\,\mathrm{nm}$) on a timescale of less than $150\,\mathrm{fs}$. These findings will allow to further study and characterize ultrafast hot carrier transport and its manipulation, e.g., by defects and impurities, not only in bismuth but also in other materials.
	
	\begin{acknowledgements}
		We gratefully acknowledge fruitful discussions with R. Merlin and A. von Hoegen. Funding by the Deutsche Forschungsgemeinschaft (DFG, German Research Foundation)
		through project B04 of Collaborative Research Center SFB 1242 “Nonequilibrium dynamics of
		condensed matter in the time domain” (Project\nobreakdash-ID~278162697) is appreciated.
		G.S. acknowledges the support of the National Science and Engineering Research Council of Canada, the Ontario Early Researcher Award program, the Canada Foundation for Innovation, and the Canada Research Chair program. The authors declare no competing financial interest.
	\end{acknowledgements}

\bibliography{main} % Produces the bibliography via BibTeX. No other commands are neccessary.
	
\end{document}